\documentclass[aps,prl,reprint,superscriptaddress,twocolumn]{revtex4}

\usepackage{amsfonts}
\usepackage{amsmath}
\usepackage[normalem]{ulem}
\usepackage{bm}
\usepackage{graphicx}
\usepackage{subfig}
\usepackage{lipsum}
\usepackage{caption}
\usepackage{array,multirow}
\usepackage[utf8x]{inputenc}
\newcommand{\ket}[1]{\vert#1\rangle}

\def\opone{\leavevmode\hbox{\small1\kern-3.8pt\normalsize1}}
\usepackage{color}

\begin{document}
	
	\title{
		Spectrally Multiplexed Hong-Ou-Mandel Interference}
	
	\author{Oriol Pietx-Casas}
	\altaffiliation{Corresponding author: O.PietxiCasas@tudelft.nl}
	\affiliation{QuTech, Delft University of Technology, 2600 GA Delft, The Netherlands}
	\affiliation{Kavli Institue of Nanoscience, Delft University of Technology, 2600 GA Delft, The Netherlands}
	%-----------------------------
	\author{Gustavo Castro do Amaral}
	\affiliation{QuTech, Delft University of Technology, 2600 GA Delft, The Netherlands}
	\affiliation{Center for Telecommunication Studies, Pontifical Catholic University of Rio de Janeiro,\\ 22451-900, Rio de Janeiro, Brazil}
	%-----------------------------
	\author{Tanmoy Chakraborty}
	\affiliation{QuTech, Delft University of Technology, 2600 GA Delft, The Netherlands}
	\affiliation{Kavli Institue of Nanoscience, Delft University of Technology, 2600 GA Delft, The Netherlands}
	%-----------------------------
	\author{Remon Berrevoets}
	\affiliation{QuTech, Delft University of Technology, 2600 GA Delft, The Netherlands}
	%-----------------------------
	\author{Thomas Middelburg}
	\affiliation{QuTech, Delft University of Technology, 2600 GA Delft, The Netherlands}
	%-----------------------------
	\author{Joshua Slater}
	\affiliation{QuTech, Delft University of Technology, 2600 GA Delft, The Netherlands}
	%-----------------------------
	\author{Wolfgang Tittel}
	\affiliation{QuTech, Delft University of Technology, 2600 GA Delft, The Netherlands}
	\affiliation{Kavli Institue of Nanoscience, Delft University of Technology, 2600 GA Delft, The Netherlands}
	\affiliation{Department of Applied Physics, University of Geneva, 1211 Geneva 4, Switzerland}
	\affiliation{Schaffhausen Institute of Technology in Geneva, Switzerland}
	
	\begin{abstract}
		We explore the suitability of a Virtually-Imaged Phased Array (VIPA) as a Spectral-to-Spatial Mode-Mapper (SSMM) for applications in quantum communication such as a quantum repeater. To this end we demonstrate spectrally-resolved two-photon ``Hong-Ou-Mandel'' (HOM) interference. Spectral sidebands are generated on a common optical carrier and weak coherent pulses are prepared in each spectral mode. The pulses are subsequently sent to a beamsplitter followed by two SSMMs and two single-photon detectors, allowing us to measure spectrally-resolved HOM interference. We show that the so-called HOM dip can be observed in the coincidence detection pattern of matching spectral modes with visibilities as high as 45\% (with the maximum being 50\%). For unmatched modes, the visibility drops significantly, as expected. Due to the similarity between HOM interference and a linear-optics Bell-state measurement (BSM), this establishes this simple optical arrangement as a candidate for the implementation of a spectrally-resolved BSM. Finally, we simulate the secret key generation rate using current and state-of-the-art parameters in a Measurement-Device Independent Quantum Key Distribution scenario and explore the trade-off between rate and complexity in the context of a spectrally-multiplexed quantum communication link.
		\end{abstract}
	
	\maketitle

\begin{figure}[ht]
    \centering
    \includegraphics[width=0.95\linewidth]{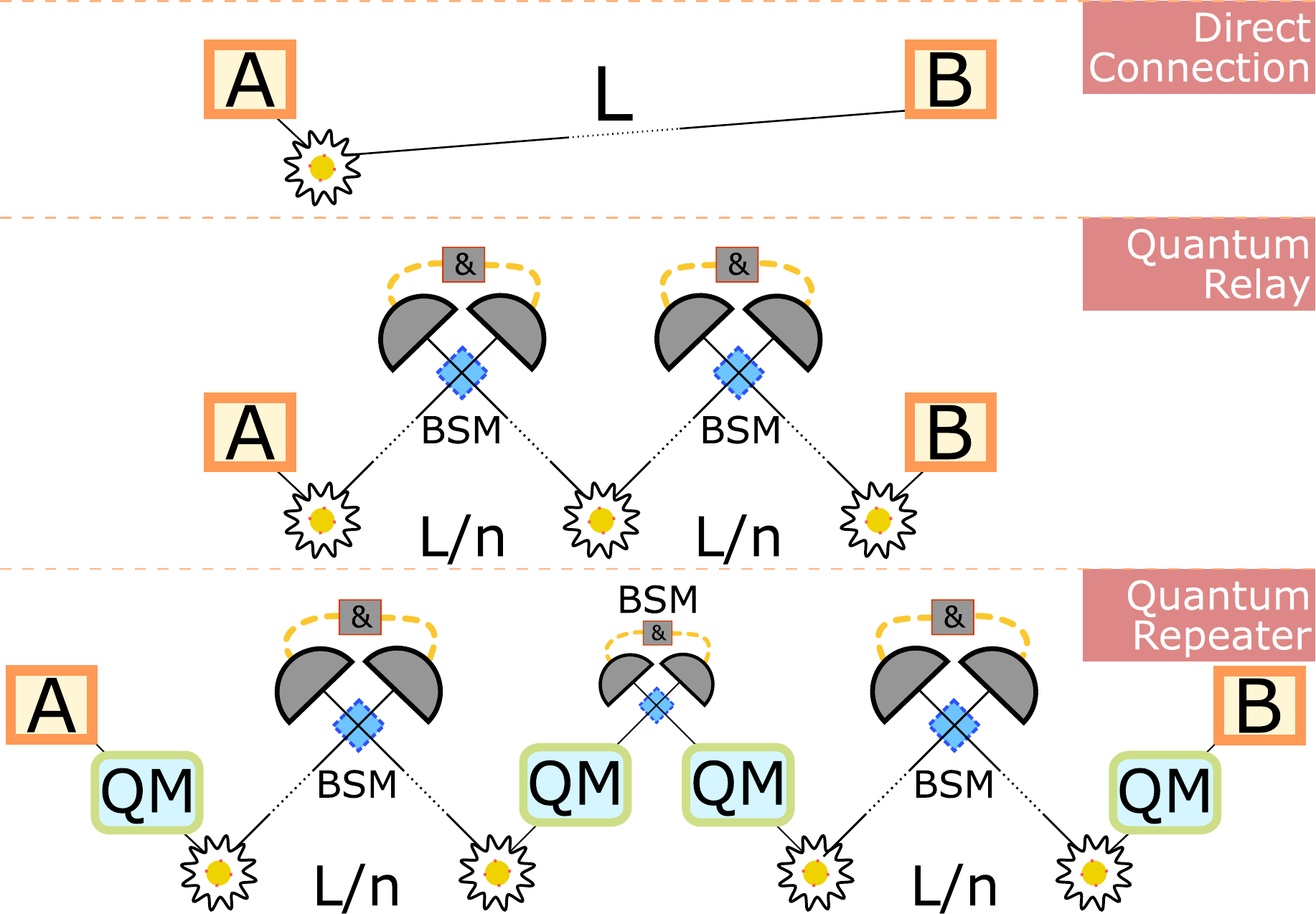}
     \caption{Three different possible configurations of a long-distance quantum communication link between end-users A and B: direct connection; the quantum relay; and the quantum repeater. The quantum repeater consists of a memory-assisted quantum relay that decouples the elementary links and allows overcoming channel loss through the use of multiplexing and feed-forward mode mapping. L is the total distance and n is the number of links.}
     \label{fig:rel_rep}
    \end{figure}

    A quantum repeater will allow long-distance connectivity of quantum nodes, ultimately enabling the establishment of the quantum internet. With a working quantum internet comes the possibility of implementing protocols such as cloud quantum computing, quantum-key-distribution (QKD) without trusted nodes, and dense coding \cite{Ekert(august1991), Benett(november1992)}. However, for these applications to run between users -- say A and B -- over arbitrary distances, a quantum repeater, in some form, becomes necessary, the simplest of which is that of a memory-assisted quantum relay: a chain of elementary links that consist of sources of entangled photon-pairs, quantum memories, a linear-optics-based Bell-state measurement (BSM), and a quantum channel. Within an elementary link, direct connection through entanglement swapping is possible by restricting the distances such that the losses in the quantum channel are constrained. To close the chain between end nodes A and B, the elementary links are interconnected by entanglement swapping, as depicted in Fig. \ref{fig:rel_rep}.
    
    Multiplexing enhances the entanglement distribution rates of a quantum repeater, and temporal multiplexing figures as the most common flavour of quantum repeater architectures, where the quantum memories must be capable of accessing the stored quantum states at any desired time. These so-called \textit{on-demand} quantum memories allow decoupling transmission through the elementary links such that the successful entanglement distribution across the entire link does not depend on simultaneous successful entanglement swapping operations within all the constituent elementary links \cite{duan2001long}. Successfully swapped entangled states are stored in the memories until either the storage time limit is reached or entanglement distribution across the neighboring elementary link has been successful.
    
    A quantum repeater making use of multiplexing is able to overcome the rates achieved by a quantum relay. The rate enhancement can be easily quantified in a simple model (see Fig. \ref{fig:rel_rep}). We start by taking, as a reference, the rate achieved by a quantum relay: since there is no rate boost for a quantum relay architecture when compared to direct communication, $R^{\text{relay}}$ can be computed, in the ideal case of unit efficiencies (including ideal BSMs), as $R^{\text{relay}} = R^{\text{source}}10^{-\alpha L}$, where $L$ is the distance between A and B in meters, $\alpha$ is the fiber's loss coefficient in dB/m, and $R^{\text{source}}$ is the maximum repetition rate achieved by the sources \cite{Sinclair(july2014)}. For a temporally-multiplexed quantum repeater, we must first define the total number of attempts $M$ that can be performed during the memory's storage time $t^{\text{store}}$, i.e., $M=t^{\text{store}}R^{\text{source}}$. It can be shown that \cite{wang2021}:
    \begin{equation}
        R^{\text{repeater}} = R^{\text{source}}\left(1-\left(1-10^{-\frac{\alpha L/n}{10}}\right)^M\right)^n,
        \label{eq:rate}
    \end{equation}
    which represents a considerable boost over $R^{\text{relay}}$ if $M$ increases.
    
    Unfortunately, some (if not all) quantum memories that allow read-out on demand introduce deadtime and hence limit the entanglement generation rate of a temporally multiplexed quantum repeater\cite{askarani2021long}. For instance, for the atomic frequency comb protocol \cite{Afzelius2009}, this problem is caused by the need for control pulses that transfer optically excited coherence onto spin coherence and, at a desired time, back. As described in \cite{askarani2021long}, no more qubits can be added to the memory after the first control pulse. This creates a bottleneck for the rate since the memories must potentially store states for long times. A spectrally-multiplexed quantum repeater can overcome this bottleneck \cite{Sinclair(july2014)}. Through a suitable design of the spectra of the photons generated by the sources (and absorbed by the memories), it is possible to perform, via spectral-multiplexing, $M$ attempts \emph{during a single temporal window}. Furthermore, an arbitrary number of temporal modes can be stored concurrently in the memory's optical coherence, contrasting the above-mentioned case of spin mapping. Under these conditions, the spectrally-multiplexed quantum memory operates with a fixed-storage time, which creates a constraint on the length of the elementary links: the heralded results of the entanglement swapping operation must be available at the memory location before its storage time elapses.  This condition can be encapsulated in mathematical form as $t^{\text{fixed}} = L/n \cdot \left(v_{\text{fiber}}\right)^{-1}$, where $v_{\text{fiber}}$ is the speed of light in the optical fiber. Extending the optical storage time of ensemble-based broadband quantum memories is an on-going effort, and recent experiments \cite{askarani2021long} have demonstrated  $t^{\text{fixed}}\!=\!100$ $\mu$s -- enough for a 20-km-long elementary link. 
    
    In the same way that temporally-multiplexed quantum memories use on-demand retrieval to reconcile entanglement in different temporal modes, spectrally-multiplexed memories require a feed-forward spectral mode-mapper to reconcile different spectral modes \cite{Sinclair(july2014)}. The information about which mode has been successful, imperative for the repeater's operation, must be determined by the remote BSM stations. They therefore require the necessary technological overhead to distinguish between different spectral modes. Since the frequency bandwidth of quantum repeaters is likely to be limited by the forthcoming quantum memories (e.g. 56~GHz in \cite{askarani2021long}), DWDM technology is unsuitable due to the limited amount of available channels within this bandwidth. Our work addresses this problem: we demonstrate high-resolution spectrally-resolved two-photon interference by making use of virtually-imaged phased arrays (VIPAs). This paves the way towards a spectrally-multiplexed quantum repeater.
    
    The Bell-State measurement is ubiquitous for long distance quantum communications---in particular for quantum repeaters---but also in functionally simpler (and thus simpler to performance benchmark) applications, such as Measurement Device-Independent QKD (MDI-QKD) \cite{Rubenok(2013)}. To investigate the performance enhancement of the proposed spectrally-multiplexed BSM, we implement a spectrally-multiplexed MDI-QKD channel and use the extracted secret key rate as a figure of merit for the BSM's performance. The proposal involves routing spectrally-multiplexed ($M>1$) attenuated laser pulses to a single beam-splitter, mapping each spectral mode to a different spatial mode (spectral-to-spatial mode mapping -- SSMM) using the VIPA, detecting each spatial mode at a different detector, and correlating the events of all possible detector combinations. We expect to observe a so-called Hong-Ou-Mandel (HOM) dip when detectors for the same spectral/spatial mode are selected.
    
    \begin{figure}[ht]
    \centering
    \includegraphics[width=0.95\linewidth]{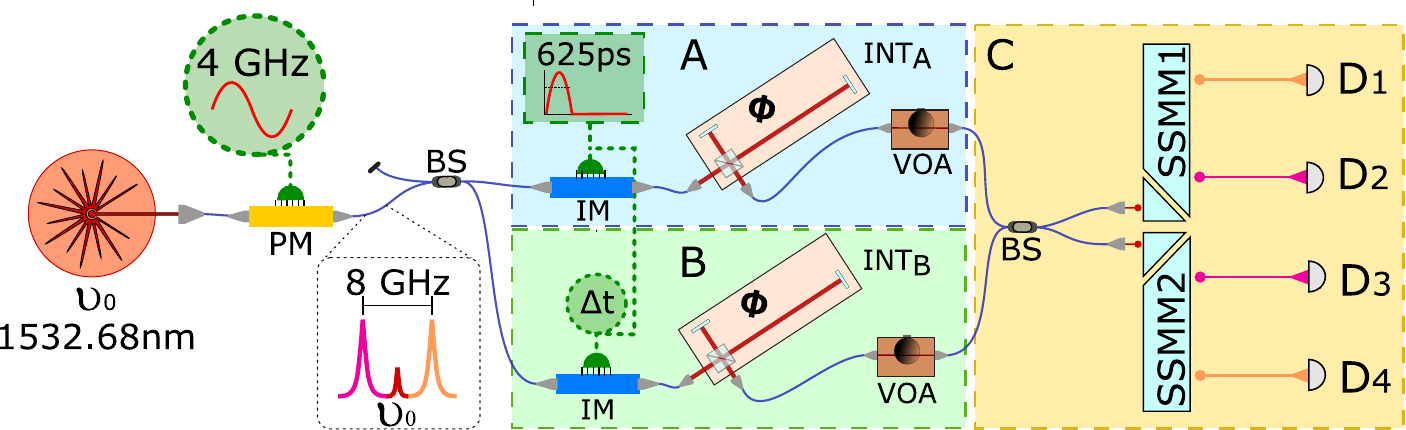}
     \caption{Experimental setup. A single beam-splitter (BS) and two SSMMs based on VIPAs allow demonstrating $M=2$ spectrally-resolved two-photon interference. PM: Phase Modulator; IM: Intensity Modulator; INT$_{\text{A,B}}$: Phase-Controlled Interferometers; VOA: Variable Optical Attenuator; $\Delta t$: Relative arrival time between pulses at the BS.}
     \label{fig:setup}
    \end{figure}
    
    The experimental setup is depicted in Fig. \ref{fig:setup}. A phase modulator driven by a 4~GHz sinusoidal signal modulates two sidebands onto a continuous-wave laser beam with central wavelength $\lambda_0=1532.68$ nm. We will refer to the red-shifted spectral mode (-4~GHz away from the central optical carrier) as spectral mode 1, and to the blue-shifted one (+4~GHz away) as spectral mode 2. The remaining optical carrier will henceforth be ignored. The light is then distributed to two preparation stations (A and B) -- located in the same laboratory -- that encode time-bin qubits using electro-optical intensity modulators (used to carve 625\textcolor{green}{-}ps-long pulses at an 80~MHz repetition rate) and unbalanced, phase-stabilized interferometers with the same optical path-length difference. The resulting pulses are furthermore attenuated to the single-photon level and transmitted to the measurement station C. See \cite{Rubenok(2013)} for more information.
    
    At the measurement station, the two beams---one from A and one from B---are combined in a symmetrical fiber-optic beam splitter, and the outputs are cast onto free-space and shaped by a cylindrical lens. Each output is subsequently directed to a different VIPA, which maps different spectral modes onto distinct spatial modes that propagate into different directions. These modes are collected into different fibres, positioned in the back focal plane of a lens with positive focal length. This arrangement constitutes an SSMM able to differentiate between different frequency modes (spectral-to-temporal solutions have also been recently demonstrated \cite{saglamyurek2016multiplexed,merkouche2021spectrally}). Note that the number of beam splitters and VIPAs remains the same as the number N of spectral/spatial modes grows. However, the number of required detectors scales as 2N, which poses a concern in terms of system scalability. Recently, arrays of single-photon detectors based on superconducting nanowires have been reported \cite{allmaras2020demonstration}, showing that the scaling issue can be alleviated by increasing the density of detectors within a single cryogenic system.
    
\begin{figure}[ht]
    \centering
    \includegraphics[width=0.95\linewidth]{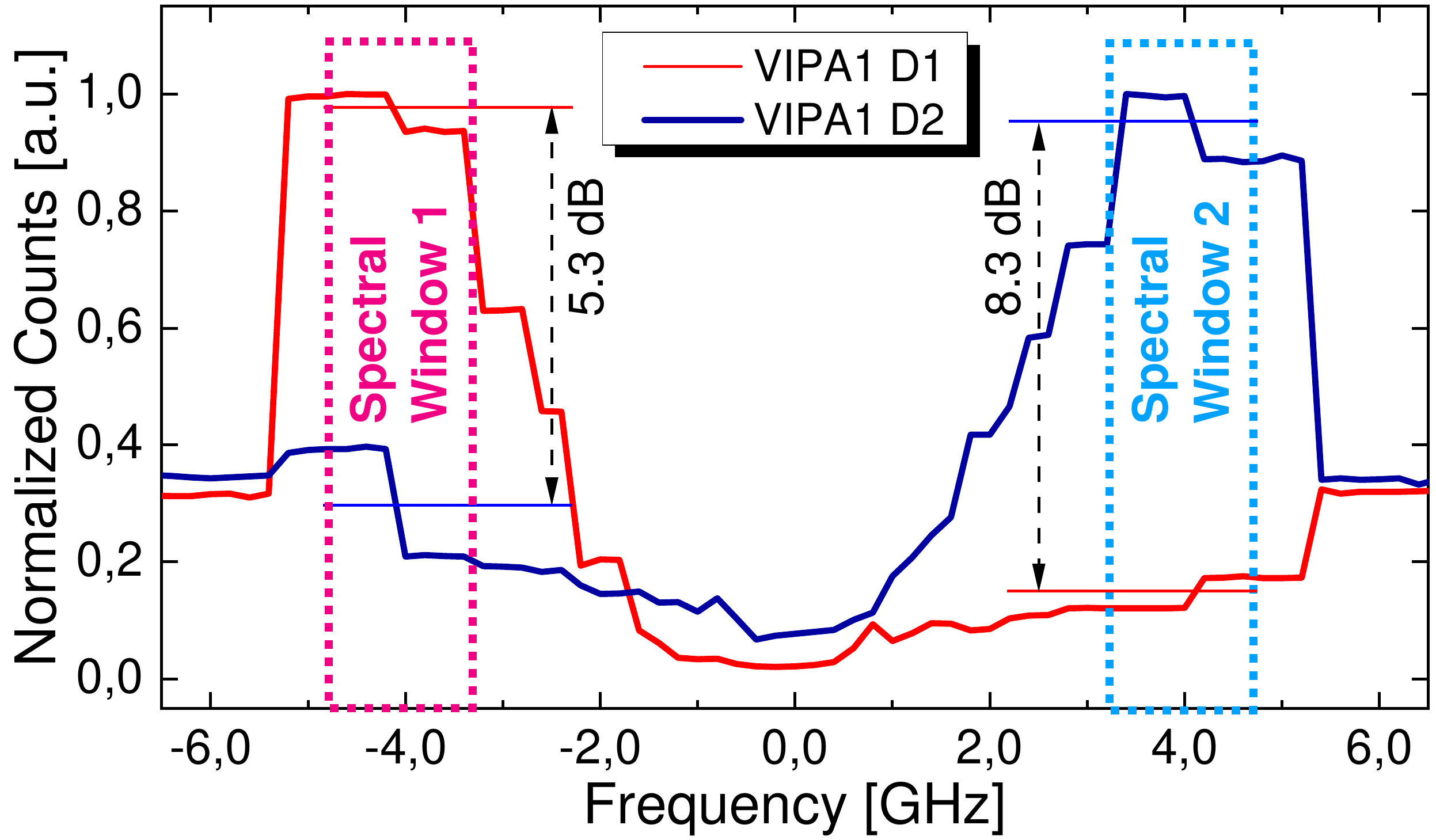}
     \caption{Frequency-response of the spectral-to-spatial mode mapper based on VIPA 1. The similar result obtained with VIPA 2 is not shown. Two distinct spectral windows can be identified at $\pm$4~GHz with respect to the center wavelength of the common laser source.}
     \label{fig:vipa-char}
    \end{figure}

    The alignment of the fibers in the focal plane of the collimating lens is critical to enable efficient coupling of the relevant frequency modes. In order to characterize the spectral resolution and the cross-talk of the assembled SSMMs, the four possible outputs -- pictorially represented in Fig. \ref{fig:setup} -- are coupled into fibers and directed to four SNSPDs. Their count rates were recorded as the frequency of the phase-modulator's modulating tone was swept in a range of [-6;6] GHz in steps of 100 MHz, yielding the frequency response of the SSMM system. In practice, the coupling can be optimized individually for each detector; however, the visibility of the two-photon interference pattern depends on the relative efficiencies of the detectors \cite{moschandreou2018experimental}. Therefore, we roughly equalized the coupling for all 4 outputs. The frequency response of SSMM 1 is presented in Fig. \ref{fig:vipa-char}, where the two distinct spectral windows separated by 8 GHz can be observed (the result obtained for SSMM 2 is similar). In these measurements, the frequency reference is taken as $\nu_0$, the center frequency of the common laser source.
    
    The current experimental setup suffers from a limited adjacent mode rejection ratio (i.e. crosstalk) and a low beamsplitter-through-VIPA-into-fiber coupling efficiency. Fig. \ref{fig:vipa-char} shows that the former is well below 10 dB, while we experimentally established the latter to be merely 5\%. This is in stark contrast with estimated values of 18 dB and 50\%, respectively \cite{xiao2005experimental}. Therefore, with an optimized optical design, the performance of the SSMM can be dramatically increased. But even under these sub-optimal conditions, it is possible to extract reasonable coincidence count rates while employing attenuated laser pulses with an average mean photon number of 0.1 for A and B. For instance, we found rates on the order of 1000 per second outside the region of indistinguishability (i.e. outside the HOM dip).
    
\begin{figure}[ht]
    \centering
     \includegraphics[width=0.95\linewidth]{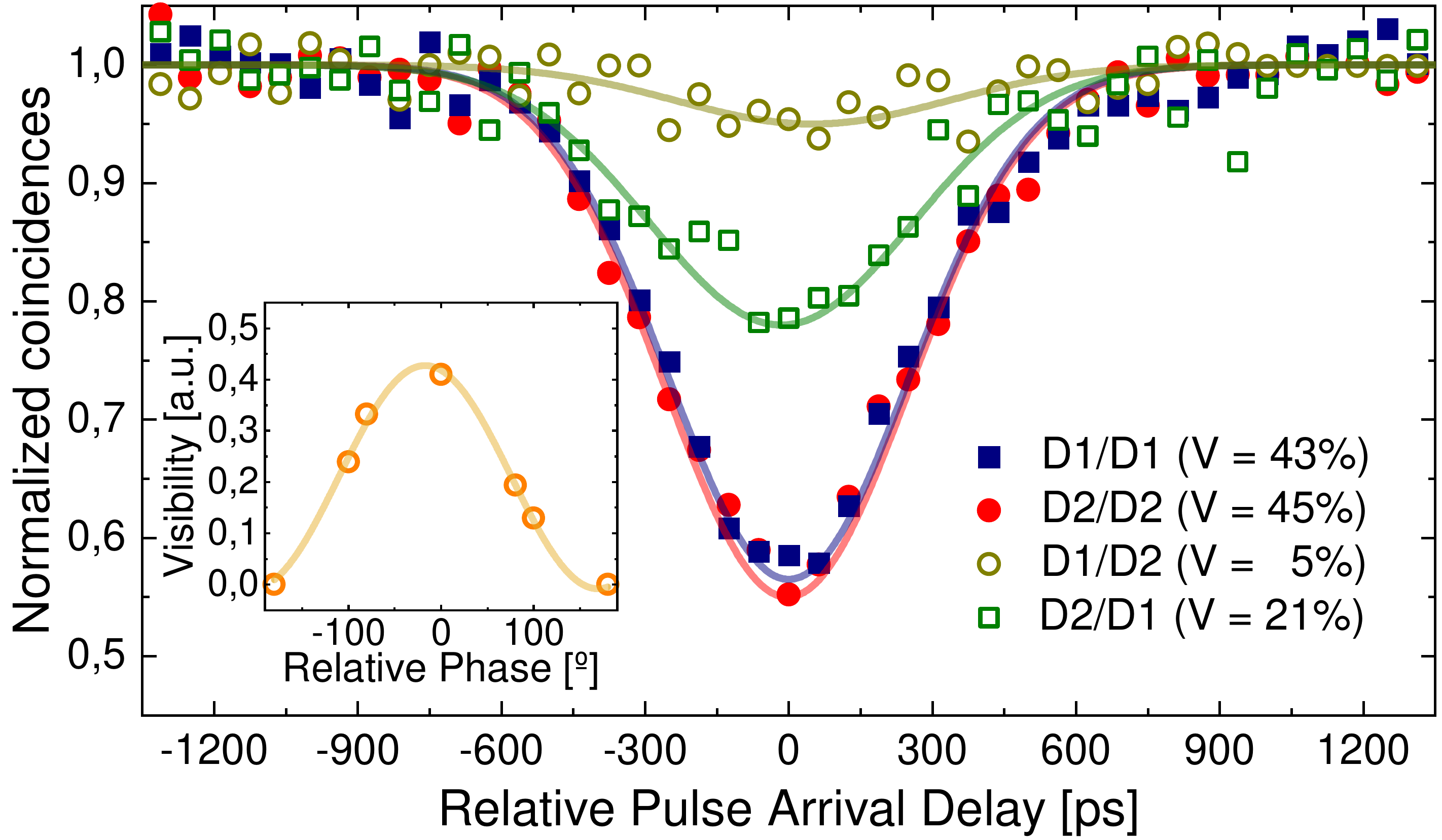}
     \caption{Coincidence counts as the relative arrival time of the attenuated laser pulses is swept for all four combinations of matched and unmatched spectral modes. Solid lines are Gaussian best fits from which the visibilities are extracted. Inset: Measured visibility of a HOM-like interference experiment where the relative phase between early and late qubit modes is varied. The solid line is a sinusoidal best fit.}
    \label{fig:HOM_dips}
    \end{figure}

    To characterise the performance of our SSMMs, we begin with a HOM dip measurement; an important figure of merit in the experiment is the visibility of the HOM dip, which is limited to 50\% for attenuated laser pulses. It is defined as $V=\tfrac{C_{\text{dist}}-C_{\text{indist}}}{C_{\text{dist}}}$, where $C_{\text{dist}}$ and $C_{\text{indist}}$ are the coincidence counts, accumulated during the same time, when the wave-packets are distinguishable and indistinguishable, respectively. In order to reach these two regions, the relative time delay of the emission of the pulses ($\Delta t$ in Fig. \ref{fig:setup}) is swept in a 3~ns range, centered at the point of maximum indistinguishability, as depicted in Fig. \ref{fig:HOM_dips}. We extract visibilities for all four possible combinations of spectral modes using time-bin qubits prepared in $\ket{late}$. These states are created by blocking the short path of the interferometers depicted in Fig. \ref{fig:setup}, and setting the mean-photon number per late temporal mode to 0.1. We found visibilities for the matching spectral modes as high as 45\%, approaching the theoretical limit of 50\% thus attesting the suitability of our setup for spectrally-resolved two-photon interference. Ideally, the visibility for unmatched spectral modes should be zero since the wave-packets should be spectrally distinguishable (i.e. feature no spectral overlap). However, as discussed above, our setup features crosstalk, resulting in visibilties of 5\% and 21\% instead (see Fig. \ref{fig:HOM_dips}).
    
    Next, we go one step further in our characterization and make use of the interferometers to create qubits of the form 
    \begin{equation}
      \begin{split}
        \ket{\psi_A} = \tfrac{1}{\sqrt{2}}\left(\ket{e}+\ket{\ell}\right)\hspace{10pt};\hspace{10pt} 
        \ket{\psi_B} = \tfrac{1}{\sqrt{2}}\left(\ket{e}+e^{i\theta}\ket{\ell}\right).
      \end{split}
    \end{equation}
    To perform a HOM-like interference measurement, we repeat the procedure used to extract the HOM visibility but look only for the coincidence events of spectral mode 1. In addition, the temporal windows within which coincidences are registered now extend over early and late temporal modes. We vary the phase difference $\theta$ between the early and late temporal modes of the qubits created in B in the range of [$-\pi$;$\pi$], and record the visibility of the HOM dip  for each value. As depicted in the inset of Fig. \ref{fig:HOM_dips}, the visibility changes from maximum to minimum as the phase causes the time-bin encoded photonic qubits to become orthogonal (this is the case if $\theta = \pm\pi$). 
    
    The final step in our analysis targets the potential performance enhancement due to the use of our SSMMs and spectral-multiplexing: how many distinct spectral modes $M$ are required to overcome the performance of a non-multiplexed system in MDI-QKD? Ideally, the answer to this question would be $M>1$; however, as we have pointed out, the inclusion of our technology induces loss and crosstalk, which significantly impacts on the system's overall performance. This likely requires $M$ to be much higher. 
    
\begin{figure}[ht]
    \centering
     \includegraphics[width=0.95\linewidth]{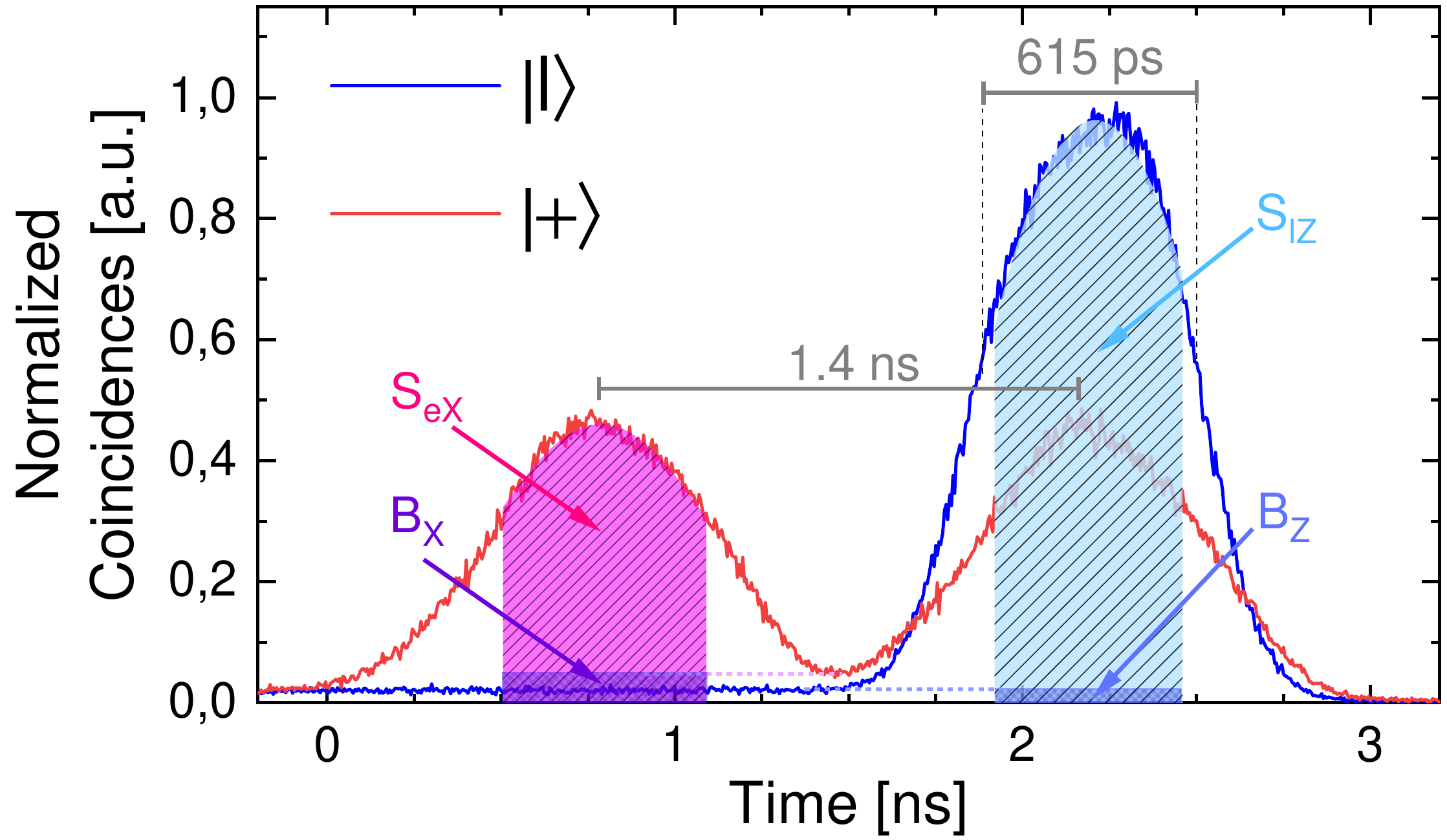}
     \caption{The set of experimentally measured parameters that constitute the inputs for the key rate estimation model. Highlighted are $S_{\ell Z}$, $B_{Z}$ and $S_{eX}$, $B_{X}$ for bases Z and X, respectively.}
    \label{fig:SandB}
    \end{figure}

    We base our calculations on \cite{yu2013three}, and the imperfections associated with the creation of a generic time-bin-encoded photonic qubit are encapsulated in the model developed in \cite{chan2014modeling}:
    \begin{equation}
        \ket{\psi} = \tfrac{1}{\sqrt{1+2b}}\Big(\sqrt{m+b}\ket{e} + e^{i\theta}\sqrt{1-m+b}\ket{l}\Big),
        \label{eq:qubit}
    \end{equation}
    with $m=S_{e}/\left(S_{e}+S_{\ell}\right)$ and $b=B/\left(S_{e}+S_{\ell}\right)$.  The experimentally determined parameters $S_{e}$, $S_{\ell}$, and $B$ are established independently for each pair of qubits belonging to a mutually unbiased basis (Z, spanned by $\ket{e}$ and $\ket{\ell}$; and X, spanned by $\ket{\pm}=(\ket{e}\pm\ket{\ell})\sqrt{2}$), as depicted in Fig. \ref{fig:SandB} for the states $\ket{\ell}$ and $\ket{+}$. Apart from $m$ and $b$, the model takes as input the observed visibility of the HOM-dip---we choose 42\% as the average between spectral mode 1 and spectral mode 2, and then allows calculating the secret key rate for the MDI-QKD channel.
    
\begin{figure}[ht]
    \centering
     \includegraphics[width=0.95\linewidth]{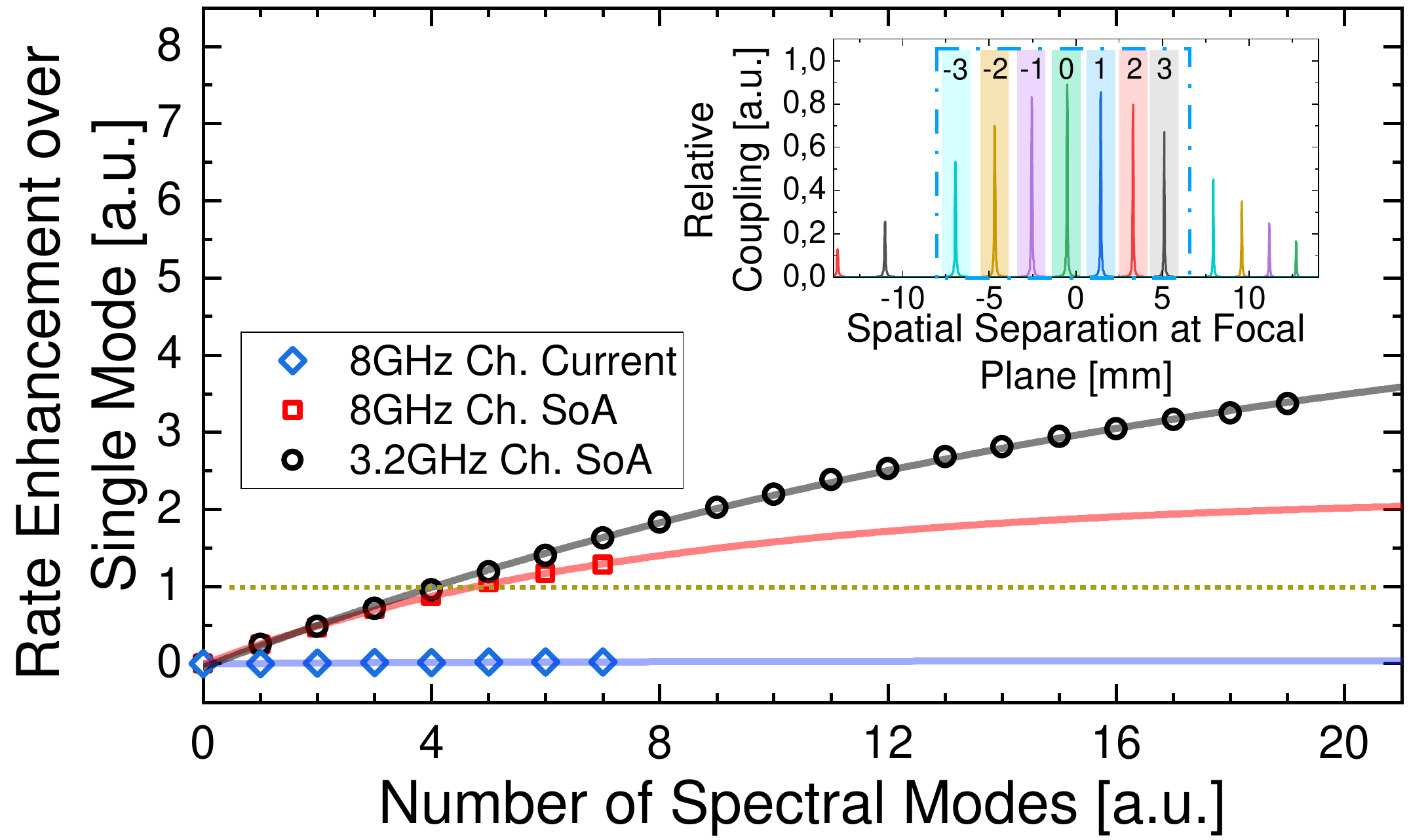}
     \caption{Enhancement of the secret key rate over the single mode case for different numbers of spectral modes assuming three different experimental conditions. SoA denotes state-of-the-art. The fits are exponential best fits that represent upper bounds. The inset showcases the decreasing coupling efficiency experienced by modes away from the VIPA's nominal operating wavelength \cite{xiao2005experimental} . Spatial separation is calculated based on a collimating lens with focal length $f=1$ meter; the blue dashed square highlights the current VIPA's spectral bandwidth.}
    \label{fig:rate}
    \end{figure}

    To assess the potential impact of the proposed spectrally-multiplexed BSM, we compute secret key rates for three different scenarios: (1) the current experimental scenario; (2) the current scenario but assuming state-of-the-art coupling efficiency of 50\% for the central mode (see the inset of Fig.~\ref{fig:rate}); and (3) a scenario where the coupling is again state-of-the-art, and the channel spacing is reduced from 8 to 3.2 GHz, i.e. limited only by the width of the spectral channels in Fig. \ref{fig:vipa-char}. These rates are compared to the regular single-mode case in which the losses from the SSMM are removed but all other parameters, (e.g. qubit descriptions as in Eq.~3, and $R^{\text{source}}$) are kept the same.
    
    The results are displayed in Fig. \ref{fig:rate}.  Our analysis includes the constraint that the VIPA in our current SSMM has a 60 GHz spectral bandwidth (matching that of the memory in \cite{askarani2021long}). Thus, assuming a mode spacing of 8 GHz, this constraints the maximum number of modes to 7. In scenario (3), this number increases to 20. Note, however, that the VIPA allows for smaller channel spacing. For instance, a spectral resolution around 600 MHz has been reported in \cite{xiao2005experimental}). In addition, VIPAs with larger spectral bandwidth can be manufactured, but with a direct impact on maximum coupling efficiency and spatial resolution; therefore, we picked the same value for the spectral bandwidth for all three scenarios. Note that the rate scales non-linearly with the number of modes, which is due to the decreasing coupling efficiency for spatial modes away from the VIPA's nominal operating wavelength -- refer to the inset of Fig. \ref{fig:rate}. 
    
    We find that the coupling loss in the current setup prevents increasing the rates, even when the number of spectral modes reaches its maximum of 7 (scenario (1)). However, we also see that the  proposed spectrally-resolved BSM outperforms a single-mode BSM for M$\geq$5 modes if the coupling is improved from its current value of 5\% to 50\% (scenario (2)). A crossover is confirmed for scenario (3), in which the mode spacing is reduced from 8 to 3.2 GHz. There, due to the larger number of possible modes, the rate enhancement grows up to a factor of 3.4. 
    
    Assuming the use of a SSMM with increased spectral range while keeping all other parameters as before allows us to extend the prediction to larger numbers of spectral modes. The observed saturation of the secret key rate with growing M indicates that it is advantageous to increase the density of spectral modes rather than to increase the total spectral window. It is interesting to observe how the time-bandwidth product (TBP) plays a role in the maximum achievable rate in the frequency-multiplexed scenario: reducing the pulse width allows increasing the rate. In turn, this widens the spectral channels, thereby reducing the maximum number that can be resolved within the total spectral bandwidth. Hence, by increasing $R^{\text{source}}$, a limit is set for $M$, which corresponds to the most efficient usage of the available spectrum and saturates Eq. \ref{eq:rate}. But note that the optimal number of spectral modes (and hence the temporal mode duration and the emission rate that maximize the TBP) depends on the use of the SSMM. For instance, in MDI-QKD \cite{Rubenok(2013)} it may be best to limit the setup to one spectral mode, while a larger number of modes is needed for a spectrally multiplexed quantum repeater \cite{Sinclair(july2014)}.
    
    In conclusion, we have proposed and experimentally demonstrated a VIPA-based spectral-to-spatial mode-mapper capable of performing spectrally-resolved two-photon interference. We characterized its properties, current limitations, and potential to implement a spectrally-resolved Bell-state measurement in the near-future, where rate enhancements in a spectrally-multiplexed quantum repeater scenario can be reached.

\noindent \textbf{Funding.} The authors acknowledge funding through the Netherlands Organization for Scientific Research (NWO), and the European Union’s Horizon 2020 Research and Innovation Program under Grant Agreement No.820445 and Project Name Quantum Internet Alliance.

% Bibliography
\bibliography{Bibliography.bib}

\end{document}